\documentclass[prl,showpacs,preprint]{revtex4}

\usepackage{amsmath,amssymb,epsfig,epsf}

\begin{document}

\title
{Electronic structure of two-dimensional hexagonal diselenides:
charge density waves and pseudogap behavior}
\author{$^1$E.Z. Kuchinskii, $^1$I.A. Nekrasov, $^{1,2}$M.V. Sadovskii}

\affiliation
{$^1$Institute for Electrophysics, Russian Academy of Sciences, Ural Branch,\\ Amundsen str. 106, Ekaterinburg, 620016, Russia\\ 
$^2$Institute for Metal Physics, Russian Academy of Sciences, Ural Branch, \\  S. Kovalevskaya str. 18, Ekaterinburg, 620219, Russia}

\begin{abstract}

We present theoretical study of electronic structure (spectral functions and
Fermi surfaces) for incommensurate pseudogap and charge density wave 
(CDW) and commensurate CDW phases of quasi two dimensional diselenides 
2H-TaSe$_2$ and 2H-NbSe$_2$.  Incommensurate pseudogap regime is described
within the scenario based on short-range order CDW fluctuations,  
considered within the static Gaussian random field model.
In contrast e.g. to high-T$_c$ cuprates layered dichalcogenides have several 
different CDW scattering vectors and electronic spectrum with two bands
at the Fermi level. To this end we present theoretical
background for the description of multiple scattering processes
within  multiple bands electronic spectrum. Thus obtained theoretical spectral 
functions and Fermi surfaces are compared with recent ARPES experimental data,
demonstrating rather good qualitative agreement.

\end{abstract}


\maketitle

\newpage

\section{Introduction}

Quasi-two-dimensional dichalcogenides TX$_2$ (T=Nb,Ta,Mo,Hf; X=S,Se)
and their different polymorhic modifications long time ago
attracted the attention of scientific community \cite{Wilson74}. This was connected with:
(i) early suggestions to look for high-T$_c$ superconductivity in
layered compounds; (ii) the discovery  of phase transitions
with formation of charge density waves (CDW) \cite{Wilson74}.
In particular, in 2H-TaSe$_2$ (2H means -- hexagonal structure with
two Ta layers in the unitcell) the second order transition into incommensurate CDW phase 
is observed at temperature Т=122.3K. At 90K there is another transition
to commensurate CDW phase \cite{Wilson74,Moncton}.
In 2H-NbSe$_2$ transition to incommensurate CDW phase happens at much lower
temperature of 33.5K \cite{Moncton} and no commensurate CDW phase is observed.

Above the temperature of incommensurate CDW transition in these systems
there might be the range of temperatures  where short-range order CDW fluctuations
with finite, but pretty large, correlation length $\xi$ may exist due to low-dimensional
nature of these systems (and in analogy with antiferromagnetic fluctuations in cuprates). This indeed
is experimentally observed in angular resolved X-ray photoemission (ARPES) experiments \cite{MSe2_1,MSe2_3,MSe2_4}.

In this paper we present band structure calculation results for
2H-TaSe$_2$ and 2H-NbSe$_2$ with analysis of possible topologies of the Fermi surfaces
upon doping, showing possibility of formation of ``bone''- like Fermi sheets.
Further we describe the details of theoretical description of multiband electronic multiple scattering on CDW
in a multiple band systems, as applied to pseudogap, incommensurate and commensurate CDW phases for both 
2H-TaSe$_2$ and 2H-NbSe$_2$. As an outcome, we obtain spectral functions and Fermi surface maps, which are compared 
with a number of recent ARPES results \cite{MSe2_1,MSe2_3}.

\section{Band structure}

The 2H-TX$_2$ layered compounds have hexagonal crystal structure
with the space group of symmetry P6$_3$/mmc with lattice
parameters for Ta system a=3.436~\AA~and c=12.7~\AA. Corresponding
Wyckoff positions are for Ta 2b (0,0,0.25) and Se 4f ($\frac{1}{3}$,$\frac{2}{3}$,0.118) \cite{Moncton}.
Formal electronic configuration of Ta is $d^1$.
To calculate electronic structure of the compound the
linearized muffin-tin orbitals method (LMTO)~\cite{LMTO} with default settings
was employed. Obtained band structure and Fermi surfaces are in good agreement
with similar LDA calculations by other authors \cite{elstruc}.
We do not present any LDA results on 2H-NbSe$_2$ since its crystal structure \cite{Moncton} and
corresponding band structure are very close to those of 2H-TaSe$_2$.

\begin{figure}
\begin{center}
\includegraphics[width=.45\textwidth]{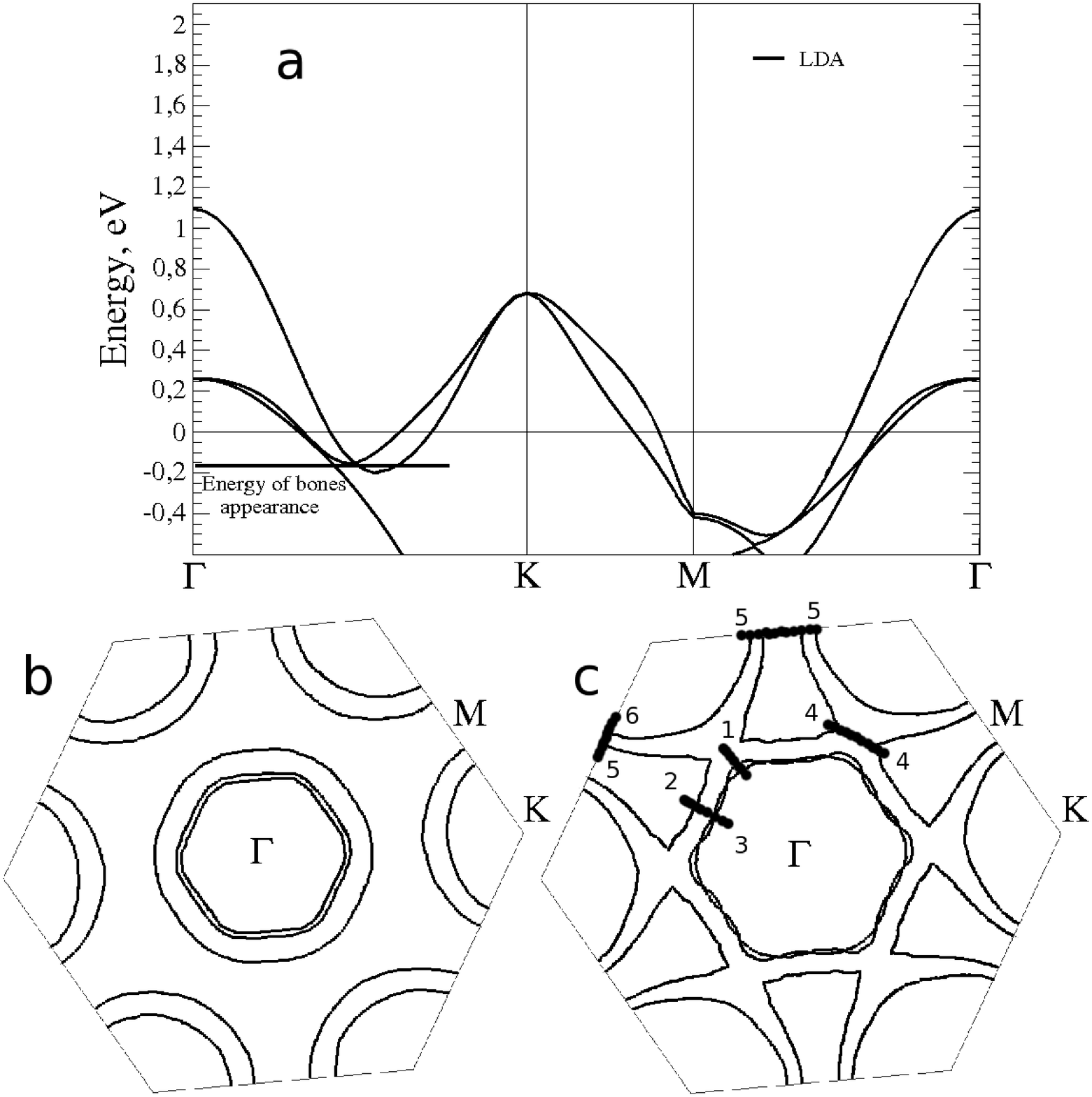}
\caption{LDA bands and Fermi surfaces for 2H-TaSe$_2$.
Panel (a) -- LDA electronic dispersions. 
Fermi level corresponds to zero.
Panel (b) -- LDA Fermi surface.
Panel (c) -- Fermi surface for shifted down Fermi level shown on panel (a)
with a short line to obtain bone-like Fermi sheets.}
\label{fig1}
\end{center}
\end{figure}

In accord with the previous works \cite{elstruc} in our LDA calculations
the Fermi level in 2H-TaSe$_2$ is crossed by two Ta-5d bands with $3z^2-r^2$ symmetry (see Fig. \ref{fig1}a),
which are well separated from other bands.

Fermi surface (FS) of 2H-TaSe$_2$ has three (in some works two \cite{elstruc})
hole-like cylinders near the $\Gamma$-point and two hole cylinders
around K-point. Our results are presented in Fig. {\ref{fig1}}b. Here we observe
three hole-like cylinders around the $\Gamma$-point.

During the recent years several ARPES studies detected the
experimental FS of 2H-TaSe$_2$. In particular, in Ref. \cite{Smith} the
electronic structure of the valence band was studied in 1T-TaS$_2$ and 2H-TaSe$_2$.
For 2H-TaSe$_2$ it was shown that along $\Gamma$-K direction
there are four crossings with the FS. Similar picture is also seen in LDA
results (Fig. {\ref{fig1}}a,b). In later ARPES works \cite{MSe2_3,Rossnagel,MSe2_1} it was observed
that FS of 2H-TaSe$_2$ has more complex topology.
Namely, along $\Gamma$-K direction there appear the
``bone''-like FS sheets. Within the LDA picture one can obtain such ``bones''
by the shift of the Fermi level down by about 0.1 eV (Fig. {\ref{fig1}}a,c).

\begin{figure}
\begin{center}
\includegraphics[width=.45\textwidth]{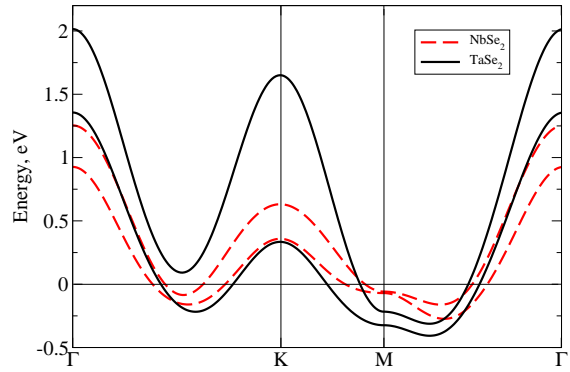}
\caption{``Experimental'' bands for 2H-TaSe$_2$ (solid line)
and 2H-NbSe$_2$ (dashed line). Fermi level corresponds to zero.}
\label{exp_band}
\end{center}
\end{figure}

To improve over simple LDA, in Fig. \ref{exp_band} we show the ``experimental'' 
bands with dispersions:
\begin{equation}
\begin{split}
\epsilon (\mathbf{k}) = t_0 &+
t_1\Bigl[2\,\mathrm{cos}\,\frac{k_x}{2}\,\mathrm{cos}\,\frac{\sqrt{3}\,k_y}{2}+\mathrm{cos}\,k_x\Bigr]+
t_2\Bigl[2\,\mathrm{cos}\,\frac{3k_x}{2}\,\mathrm{cos}\,\frac{\sqrt{3}\,k_y}{2}+\mathrm{cos}\,\sqrt{3}\,k_y\Bigr] \\&+
t_3\Bigl[\,2\,\mathrm{cos}\,k_x\,\mathrm{cos}\,\sqrt{3}\,k_y+\mathrm{cos}\,2k_x\Bigr]+
t_4\Bigl[2\,\mathrm{cos}\,3k_x\,\mathrm{cos}\,\sqrt{3}\,k_y+\mathrm{cos}\,2\sqrt{3}\,k_y\Bigr],
\end{split}
\label{spectr}
\end{equation}
with hopping integrals $t_i$ obtained from the fit to experimental Fermi 
surfaces \cite{MSe2_2}. Corresponding values of $t_i$ (in eV) for Ta system
are: for the band forming barrels around $\Gamma$ and K points
$t_0$=-0.027, $t_1$=0.199, $t_2$=0.221, $t_3$=0.028, $t_4$=0.013, 
for the band forming ``bones'' $t_0$=0.407, $t_1$=0.114, $t_2$=0.444, 
$t_3$=-0.033, $t_4$=0.011. For Nb system: for the band forming smaller cylinders
$t_0$=0.0003, $t_1$=0.0824, $t_2$=0.1667, $t_3$=0.0438, $t_4$=0.0158, while for
the band forming larger cylinders
$t_0$=0.1731, $t_1$=0.1014, $t_2$=0.2268, $t_3$=0.037, $t_4$=-0.0048.
These bands are used in further calculations below.

\section{Electronic scattering on CDW}
\subsection{Commensurate CDW phase}
\label{comCDW}

\begin{figure}
\includegraphics[clip=true,width=0.4\textwidth]{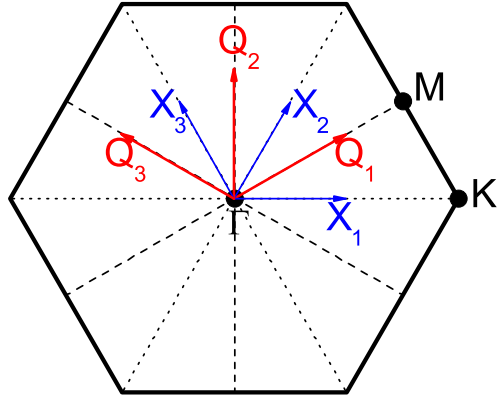}
\caption{Schematic picture of the first Brillouin zone for hexagonal lattice
with characteristic CDW vectors:
${\bf Q}=\frac{2}{3}\Gamma M$ 
(${\bf Q}_1$, ${\bf Q}_2$, ${\bf Q}_3$) -- commensurate CDW vectors. ${\bf X}=\frac{1}{2}\Gamma K$ 
(${\bf X}_1$, ${\bf X}_2$, ${\bf X}_3)$) -- vectors after two scattering on ${\bf Q}$, 
which also have significant Lindhardt function maxima.\cite{MSe2_1}}
\label{QX} 
\end{figure}

Consider the schematic picture of the first Brillouin zone for two-dimensional
hexagonal lattice, shown in Fig. \ref{QX}.
In hexagonal structures under study, the commensurate CDW vector is  
${\bf Q}=\frac{2}{3}\Gamma M$ that corresponds
to tripling the lattice period. Scattering an electron by this commensurate CDW vector
returns electron back to equivalent point after triple scattering:
$\epsilon ({\bf k}+3{\bf Q})=\epsilon ({\bf k})$. Moreover for hexagonal structures there are
in fact six equivalent  scattering vectors: 
${\bf Q}_1=(\frac{2}{3},\frac{2}{3\sqrt{3}})\pi $, ${\bf Q}_2=(\frac{2}{3},\frac{2}{3\sqrt{3}})\pi $, 
${\bf Q}_3=(-\frac{2}{3},\frac{2}{3\sqrt{3}})\pi $, and $\bar {\bf Q}_l=-{\bf Q}_l$ $(l=1,2,3)$. 
Maxima of Lindhardt function, calculated in Refs. \cite{MSe2_1,MSe2_2},
are observed on these vectors ${\bf Q}$. In addition
Lindhardt function shows pronounced maxima \cite{MSe2_1,MSe2_2} for vectors 
${\bf X}=\frac{1}{2}\Gamma K$ (${\bf X}_1=(\frac{2}{3},0)\pi $, ${\bf X}_2=(\frac{1}{3},\frac{1}{\sqrt{3}})\pi $, 
${\bf X}_3=(-\frac{1}{3},\frac{1}{\sqrt{3}})\pi $ and $\bar {\bf X}_l=-{\bf X}_l$ $(l=1,2,3)$),
which appear as sums of scattering vectors ${\bf Q}$ (see Table I of momenta summation).

Thus an electron with momentum $\bf k$ is scattered by any of thirteen different momenta (see Table I):
{\bf 0} -- preserving its initial momentum ${\bf k}$; ${\bf Q}$ (${\bf Q}_l$ and $\bar {\bf Q}_l$); 
${\bf X}$ (${\bf X}_l$ and $\bar {\bf X}_l$).
Thus for one band case to find the diagonal Green's function of an electron $G({\bf k},{\bf k})$ and twelve off-diagonal
($G({\bf k}\pm {\bf Q}_l,{\bf k})$ and ($G({\bf k}\pm {\bf X}_l,{\bf k})$) single-electron 
Green's functions we have to solve the system of thirteen linear equations (\ref{system}) (see Appendix). 
Such an approach can be generalized for a multiple band case with simplifying assumption \cite{PG_FeAs}
that intra- and interband CDW scattering amplitudes are just the same (see Appendix) 
Solving these equations we can finally find the diagonal Green function $G^{ij}({\bf k},{\bf k})$ ($i,j=1,2$ -- band indices)
and corresponding spectral function:
\begin{equation}
A(E,{\bf k})=-\frac{1}{\pi}Im\sum_{i}G^{ii}({\bf k},{\bf k})
\label{SpDen}
\end{equation}
determining the effective electron dispersion.

\begin{table}
\label{Tab1}
\caption {Table of scattering vectors summation}  
\begin{tabular}{|c||c|c|c|c|c|c|}
\hline
&  ${\bf Q}_1$ & ${\bf Q}_2$  & ${\bf Q}_3$ & $\bar {\bf Q}_1$ & $\bar {\bf Q}_2$ & $\bar {\bf Q}_3$ \\  
\hline \hline
${\bf Q}_1$ &  $\bar {\bf Q}_1$ & ${\bf X}_2$  & ${\bf Q}_2$ & 0 & $\bar {\bf Q}_3$ & $\bar {\bf X}_1$ \\ 
\hline 
${\bf Q}_2$ &  ${\bf  X}_2$ & $\bar {\bf Q}_2$  & ${\bf X}_3$ & ${\bf Q}_3$ & 0 & ${\bf Q}_1$ \\ 
\hline
${\bf Q}_3$ &  ${\bf Q}_2$ & ${\bf X}_3$  & $\bar {\bf Q}_3$ & $\bar {\bf X}_3$ & $\bar {\bf Q}_1$ & 0 \\
\hline
$\bar {\bf Q}_1$ & 0 &  ${\bf Q}_3$ & $\bar {\bf X}_1$  & ${\bf Q}_1$ & $\bar {\bf X}_2$ & $\bar {\bf Q}_2$ \\
\hline
$\bar {\bf Q}_2$ & $\bar {\bf Q}_3$ & 0 &  $\bar {\bf Q}_1$ & $\bar {\bf X}_2$  & ${\bf Q}_2$ & $\bar {\bf X}_3$ \\
\hline
$\bar {\bf Q}_3$ & ${\bf X}_1$ &  ${\bf Q}_1$ & 0 & $\bar {\bf Q}_2$  & $\bar {\bf X}_3$ & ${\bf Q}_3$ \\
\hline
\end{tabular}
\hspace{1cm}
\begin{tabular}{|c||c|c|c|c|c|c|}
\hline
&  ${\bf Q}_1$ & ${\bf Q}_2$  & ${\bf Q}_3$ & $\bar {\bf Q}_1$ & $\bar {\bf Q}_2$ & $\bar {\bf Q}_3$ \\  
\hline \hline
${\bf X}_1$ & $\bar {\bf Q}_2$ & $\bar {\bf Q}_1$  & ${\bf Q}_1$ & $\bar {\bf Q}_3$ & ${\bf Q}_3$ & ${\bf Q}_2$ \\ 
\hline 
${\bf X}_2$ & ${\bf Q}_3$ & $\bar {\bf Q}_3$  & $\bar {\bf Q}_2$ & ${\bf Q}_2$ & ${\bf Q}_1$ & $\bar {\bf Q}_1$ \\ 
\hline
${\bf X}_3$ & $\bar {\bf Q}_2$ & $\bar {\bf Q}_1$  & ${\bf Q}_1$ & $\bar {\bf Q}_3$ & ${\bf Q}_3$ & ${\bf Q}_2$ \\
\hline
$\bar {\bf X}_1$ & ${\bf Q}_3$ & $\bar {\bf Q}_3$  & $\bar {\bf Q}_2$ & ${\bf Q}_2$ & ${\bf Q}_1$ & $\bar {\bf Q}_1$ \\ 
\hline
$\bar {\bf X}_2$ & $\bar {\bf Q}_2$ & $\bar {\bf Q}_1$  & ${\bf Q}_1$ & $\bar {\bf Q}_3$ & ${\bf Q}_3$ & ${\bf Q}_2$ \\
\hline
$\bar {\bf X}_3$ & ${\bf Q}_3$ & $\bar {\bf Q}_3$  & $\bar {\bf Q}_2$ & ${\bf Q}_2$ & ${\bf Q}_1$ & $\bar {\bf Q}_1$ \\ 
\hline
\end{tabular}
\end{table}

\subsection{Incommensurate CDW phase}
\label{incCDW}

\begin{figure}
\includegraphics[clip=true,width=0.6\textwidth]{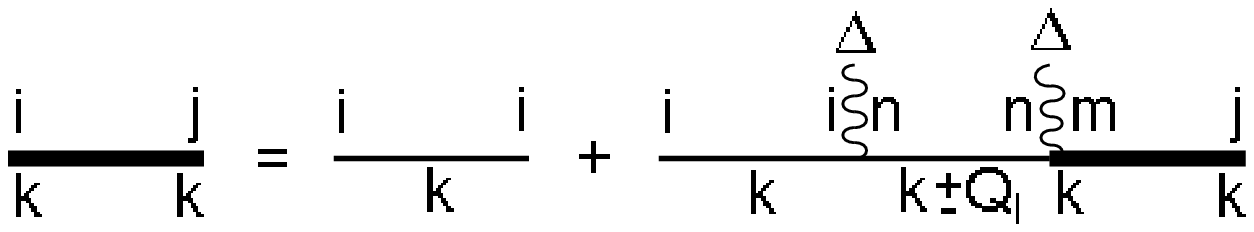}
\caption{Diagrammatic representation of diagonal Green function within two-wave approximation
for electron scattering on CDW.}
\label{diag2w} 
\end{figure}

As was pointed above at temperature $T=90$ K  2H-TaSe$_2$ (and 2H-Nb$Se_2$ at 33.5K) undergoes 
phase transition into incommensurate CDW phase with scattering vector ${\bf Q}\sim 0.58-0.6\Gamma M$.
Similar to the commensurate case discussed above, this vector corresponds to six independent scattering vectors
${\bf Q}_l, \bar {\bf Q}_l$ $l=1,2,3$. 
Let us consider single scattering of an electron with momenta $\bf k$ near the FS
by vector ${\bf Q}$(${\bf Q}_l, \bar {\bf Q}_l$). For general values of $\bf k$, such scattering act moves
an electron quite far away from the FS,  the only exception is an electron in the vicinity of the ``hot-spots''
where $\epsilon ({\bf k}+{\bf Q})=\epsilon ({\bf k})$).
Among multiple scattering processes most probable will be successive scattering processes
by vectors ${\bf Q}_l$ and $\bar {\bf Q}_l$ since in this case the scattered electron 
is back again to initial point with momenta $\bf k$ close to the Fermi surface.
To this end further we will work in the so called two-wave approximation, when
scattering act consists of two successive scattering processes by vectors ${\bf Q}_l$ and $\bar {\bf Q}_l$.
Assuming that scattering amplitude is the same for intra- and interband transitions,
for diagonal Green function's we obtain (corresponding diagrammatic representation
see in Fig. \ref{diag2w}):
\begin{equation}
G^{ij}({\bf k},{\bf k})=g^i({\bf k})\delta_{ij}+g^i({\bf k})\Sigma\sum_{m}G^{mj}({\bf k},{\bf k}),
\label{Gij}
\end{equation}
where $\Sigma = \Delta^2\sum_{jl}(g^j({\bf k}+{\bf Q}_l)+g^j({\bf k}-{\bf Q}_l))$ and  $g^j({\bf k})=\frac{1}{E-\epsilon_j({\bf k})+i\delta}$ 
is  bare retarded Green's function for the $n$-th band. Summing Eq. (\ref{Gij}) over $i$, one can get:
\begin{equation}
\sum_{i}G^{ij}({\bf k},{\bf k})=\frac{g^j({\bf k})}{1-\Sigma \sum_{i}g^i({\bf k})}.
\label{sumGij}
\end{equation}
Then using Eq. (\ref{Gij}) again we obtain:
\begin{equation}
G^{ij}({\bf k},{\bf k})=g^i({\bf k})\delta_{ij}+\frac{g^i({\bf k})\Sigma g^j({\bf k})}
{1-\Sigma \sum_{i}g^i({\bf k})},
\label{Gij_inc}
\end{equation}
which can provide us with the spectral function (\ref{SpDen}) 
for the case of incommensurate CDW scattering.

\subsection{CDW pseudogap fluctuations}
\label{PG_CDW}

\begin{figure}
\includegraphics[clip=true,width=0.5\textwidth]{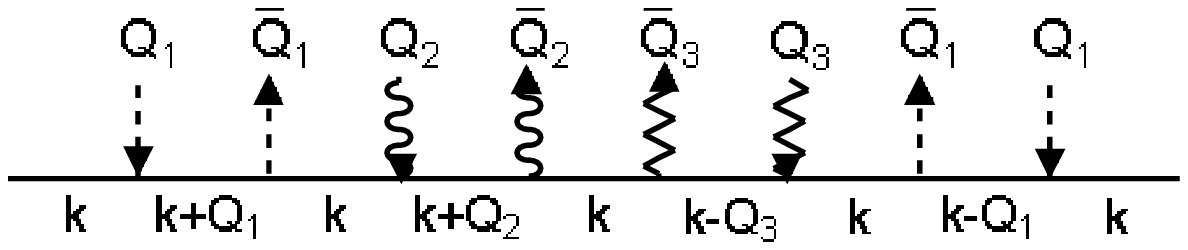}
\caption{Example of a diagram with multiple scattering on CDW,
where dashed, wavy and zig-zag lines incoming lines corresponds
to scattering on ${\bf Q}_1$, ${\bf Q}_2$, ${\bf Q}_3$, and corresponding outgoing lines --  
$\bar {\bf Q}_1$, $\bar {\bf Q}_2$, $\bar {\bf Q}_3$.}
\label{diag_2w} 
\end{figure}

Above the temperature of incommensurate CDW transition there is no long-range charge
ordering,  but due to low-dimensionality of the system there are rather well developed short-range order CDW fluctuations,
with finite correlation length $\xi$ and characteristic wave-vector $\bf Q$ which becomes rather
quick commensurate with ${\bf Q}=\frac{2}{3}\Gamma M$ \cite{MSe2_3} as temperature lowers.
In analogy to incommensurate electronic CDW scattering we will employ
two-wave approximation with pair of vectors (${\bf Q}_l, \bar {\bf Q}_l$).
Diagrammatically such scattering processes are show in Fig. \ref{diag_2w},
where three types of interaction lines correspond to three characteristic transfer momenta $l=1,2,3$.

Let us assume fluctuations  Gaussian. Then averaging over such
fluctuations corresponds to all possible interconnections of incoming and outgoing interaction lines
of the same type \cite{Sad00,KS99,Sch}, producing appropriate effective interactions, assumed to be of
the form discussed in these works.
For the case of high enough temperatures one can neglect dynamics of fluctuations and
average over static random field of Gaussian pseudogap fluctuations \cite{Sad00,KS99,Sch}.

Let us mention that the number of different diagrams is defined by product of number
of ways to interconnect vertices of type 1, 2 and 3. Since only outgoing and incoming
lines of each type can be connected,  combinatorics corresponds to incommensurate case \cite{Sad00}.
Following Refs. \cite{Sad00,KS99,Sch} we use the basic property of the diagrams of this model: any
diagram with crossing interaction lines is equal to some noncrossing diagram of the same order.
Thus only noncrossing diagrams can be considered, while contributions of {\em all}
diagrams can be accounted by combinatorial prefactors. And for each type of interaction lines (1,2,3) we will have 
its own incommensurate combinatorial prefactors, same as in Refs.  \cite{Sad00,KS99}.

\subsubsection{Recurrent procedure for Green's function: single band case}

\begin{figure}
\includegraphics[clip=true,width=0.8\textwidth]{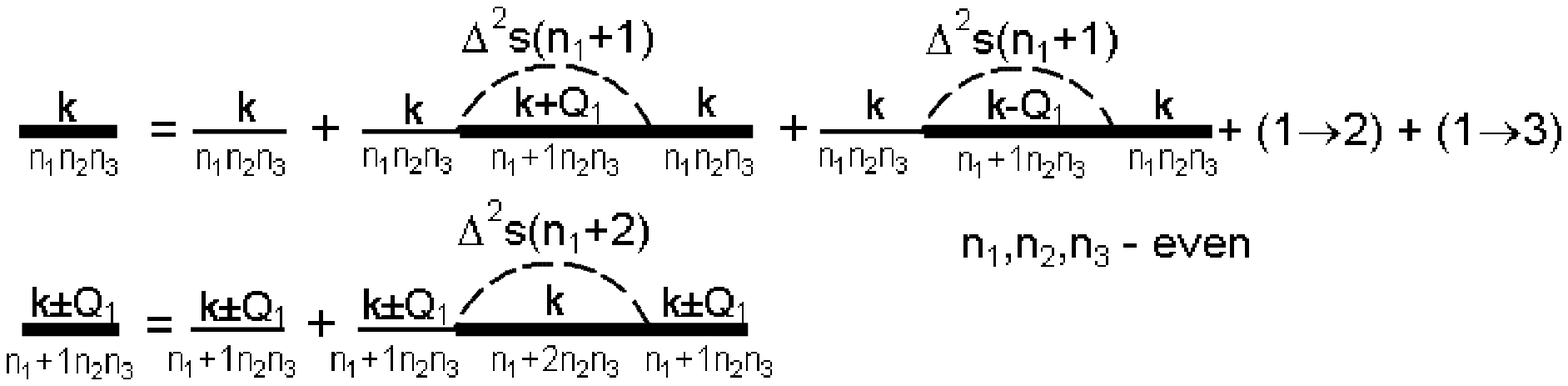}
\caption{Diagrammatic representation of Green function within single band pseudogap model
for two dimensional hexagonal systems.  (1$\to$2) denotes the two last terms, where
substitutions $Q_1\to Q_2$ and $n_1+1\to n_2+1$ should be done.}
\label{diag_1z} 
\end{figure}

Within some straightforward generalization of the approach of Refs. \cite{Sad00,KS99},
for single band case one-electron Green's function can be obtained via
recurrent procedure, which is shown diagrammatically in Fig. \ref{diag_1z}. Here
 $n_l$ is the  number of interaction lines of type $l$ surrounding the ``bare'' electron line.
Analytically this procedure can be written as: 
\begin{equation}
G_{n_1,n_2,n_3}^{-1}({\bf k})=g_{n_1,n_2,n_3}^{-1}({\bf k})-\Sigma_{n_1+1,n_2,n_3}-\Sigma_{n_1,n_2+1,n_3}-\Sigma_{n_1,n_2,n_3+1},
\label{Gn_ev}
\end{equation}
where $n_1$, $n_2$, $n_3$ -- are even and
\begin{equation}
\Sigma_{n_1+1,n_2,n_3}=
\Delta^2s(n_1+1)[G_{n_1+1,n_2,n_3}({\bf k}+{\bf Q_1})+G_{n_1+1,n_2,n_3}({\bf k}-{\bf Q_1})].
\label{Sign_od}
\end{equation}
The other self-energies $\Sigma$ in (\ref{Gn_ev}) can be found similarly to (\ref{Sign_od}), but 
$n_2$ or $n_3$ should be increased by one and vectors ${\bf Q_2}$ or ${\bf Q_3}$
should be added (subtracted) to (from) ${\bf k}$, while
\begin{equation}
G_{n_1+1,n_2,n_3}^{-1}({\bf k}\pm {\bf Q_1})=g_{n_1+1,n_2,n_3}^{-1}({\bf k}\pm {\bf Q_1})-\Sigma_{n_1+2,n_2,n_3},
\label{Gn_od}
\end{equation}
and
\begin{equation}
\Sigma_{n_1+2,n_2,n_3}=
\Delta^2s(n_1+2)G_{n_1+2,n_2,n_3}({\bf k}).
\label{Sign_ev}
\end{equation}
Here 
\begin{equation}
g_{n_1,n_2,n_3}({\bf k})=\frac{1}{E-\epsilon ({\bf k})+inv({\bf k})\kappa}, 
\label{gkapp}
\end{equation}
and $\kappa = 1/\xi$ is the inverse correlation length of pseudogap fluctuations, $n=n_1+n_2+n_3$, $v({\bf k})=|v_x({\bf k})+v_y({\bf k})|$, 
$v_{x,y}({\bf k})=\frac{\partial \epsilon ({\bf k})}{\partial k_{x,y}}$ are projections of quasiparticle velocities. 

For the case of  incommensurate fluctuations, combinatorial prefactors are:  
\begin{equation}
s(n)=\left\{\begin{array}{cc}
\frac{n+1}{2} & \mbox{for odd $k$} \\
\frac{n}{2} & \mbox{for even $k$}
\end{array} \right.
\label{vinc}
\end{equation}

This recurrent procedure is applied in analogy with refs. \cite{Sad00,KS99}. 
As a first step, one takes large enough $n=n_1+n_2+n_3$, for example even, and assume
that $all$ Green functions $G_{n_1,n_2,n_3}$ with even $n_1$, $n_2$, $n_3$, 
such that $n=n_1+n_2+n_3$ are equal to zero. Then from  (\ref{Sign_ev}) one can find
that all $\Sigma_{n_1,n_2,n_3}$ for the same indices are equal to zero too.
Then using recurrent procedure, one can get $all$ new values for
$G_{n_1,n_2,n_3}$ with even $n_1$, $n_2$, $n_3$, such that $n_1+n_2+n_3=n-2$, 
and repeat the recurrent procedure until one obtains physical Green function:
\begin{equation}
G({\bf k})=G_{0,0,0}({\bf k}).
\label{Gphys}
\end{equation}

\subsubsection{Multiple bands pseudogap model for quasi two dimensional hexagonal structures.}

\begin{figure}
\includegraphics[clip=true,width=0.8\textwidth]{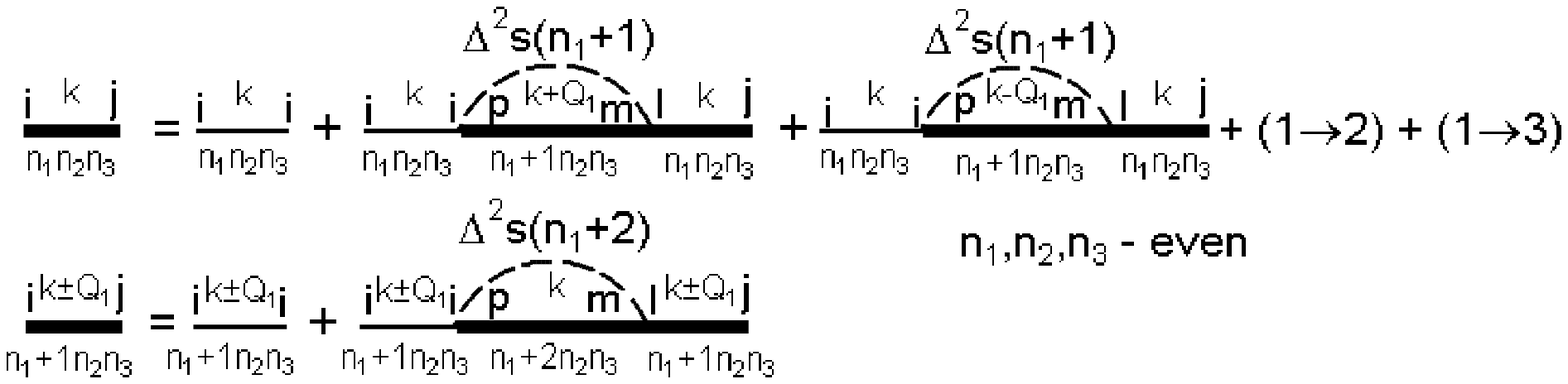}
\caption{Diagrammatic representation of Green function within multiband pseudogap model
for two dimensional hexagonal systems.}
\label{diag_nz} 
\end{figure}

In hexagonal diselenides TaSe$_2$ and  NbSe$_2$,  as we have seen, the  Fermi level is crossed by two bands.
Thus, our recurrent procedure should be generalized for multiple bands.
We will follow  Ref. \cite{PG_FeAs}, devoted  to description of possible pseudogap behavior in iron based superconductors,
and assume that intra and interband pseudogap scattering amplitudes are identical. This simplifies further analysis and
the recurrent procedure for diagonal elements of the general matrix (over band indices) Green's function
$G^{ij}$ can be drawn diagrammatically as in Fig. \ref{diag_nz}.
For our two band model each of the band indices run over two possible values,
and there is summation over all possible values of indices $p$, $m$, $l$ in the vertices (Fig. \ref{diag_nz}).
Thus, the self-energy in these diagrams has no dependence
on band indices at all and we can obtain the recurrent procedure
for $G_{n_1,n_2,n_3}=\sum_{i,j}G_{n_1,n_2,n_3}^{ij}$,  which is identical to 
Eqs. (\ref{Gn_ev})-(\ref{Sign_ev}) as in single band case, with only replacement:
\begin{equation}
g_{n_1,n_2,n_3}({\bf k})=\sum_{i}g^j_{n_1,n_2,n_3}({\bf k})=
\sum_{j}\frac{1}{E-\epsilon_j({\bf k})+inv_j({\bf k})\kappa}\qquad n=n_1+n_2+n_3.
\label{G0_nz}
\end{equation}
and at the end of the procedure we define the physical matrix Green function as:
\begin{equation}
G^{ij}({\bf k})=g^i_{0,0,0}({\bf k})\delta_{ij}+\frac{g^i_{0,0,0}({\bf k})\Sigma g^j_{0,0,0}({\bf k})}
{1-\Sigma g_{0,0,0}({\bf k})},
\label{Gij_pg}
\end{equation}
where $\Sigma=\Sigma_{1,0,0}+\Sigma_{0,1,0}+\Sigma_{0,0,1}$.
This  Green's  function allows us to find the spectral function (\ref{SpDen}) in the presence of CDW pseudogap fluctuations.

\section{Results and discussion}

In our calculations we used rather typical estimate of CDW potential
$\Delta=$0.05 eV, and for correlations length of pseudogap fluctuations we
assumed the value of $\xi$=10$a$ (where $a$ is the lattice spacing).
To mimic for the experimental ARPES resolution we broadened our spectral functions
with Lorentzian of width $\gamma = 0.03 eV$, practically it means that we made substitution 
$E\to E+i\gamma$ during all calculations.

\begin{figure}
\begin{center}
\includegraphics[width=1.\textwidth]{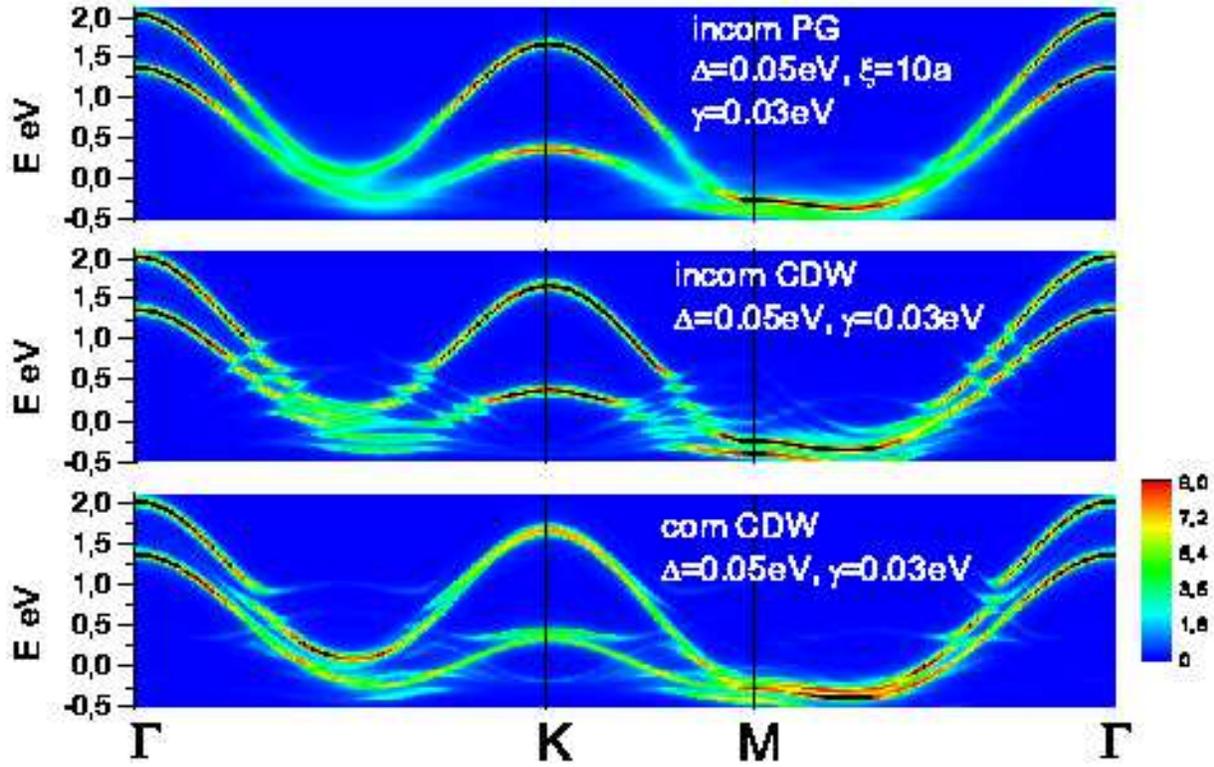}
\caption{Spectral functions of 2H-TaSe$_2$. Upper panel --
incommensurate pseudogap phase, middle panel -- incommensurate CDW phase,
lower panel -- commensurate CDW phase.}
\label{sp_full}
\end{center}
\end{figure}

In Fig. \ref{sp_full} we show spectral function maps along high symmetry directions with $k_z$=0 for 2H-TaSe$_2$.
Upper panel shows spectral function map for incommensurate pseudogap phase
obtained within our pseudogap model. In general it reminds bare ``experimental'' dispersions
plotted on Fig. \ref{exp_band}. However, here we see some {\em additional}
broadening of initial spectra. These broadened regions of spectral functions with lower intensity
represent  regions  pseudogap formation.
Why do we speak about regions? In contrast to cuprates \cite{MS} where we have
finite and rather small number of ``hot-spots'' here we have almost infinite number
of ``hot-spots'' and it is the interplay between all of them, which leads to the formation of such regions.
But still dispersions here does not have any obvious discontinuities.

Middle panel of Fig. \ref{sp_full} displays the case of incommensurate 
CDW phase. Now we see that regions previously covered with pseudogap
have clear discontinuities  --  gaps and also many shadow bands.
When we transfer further to commensurate CDW phase (lower panel of Fig.\ref{sp_full})
those  gaps become even stronger and we can see much more pronounced shadow bands.

\begin{figure}
\begin{center}
\includegraphics[width=1.\textwidth]{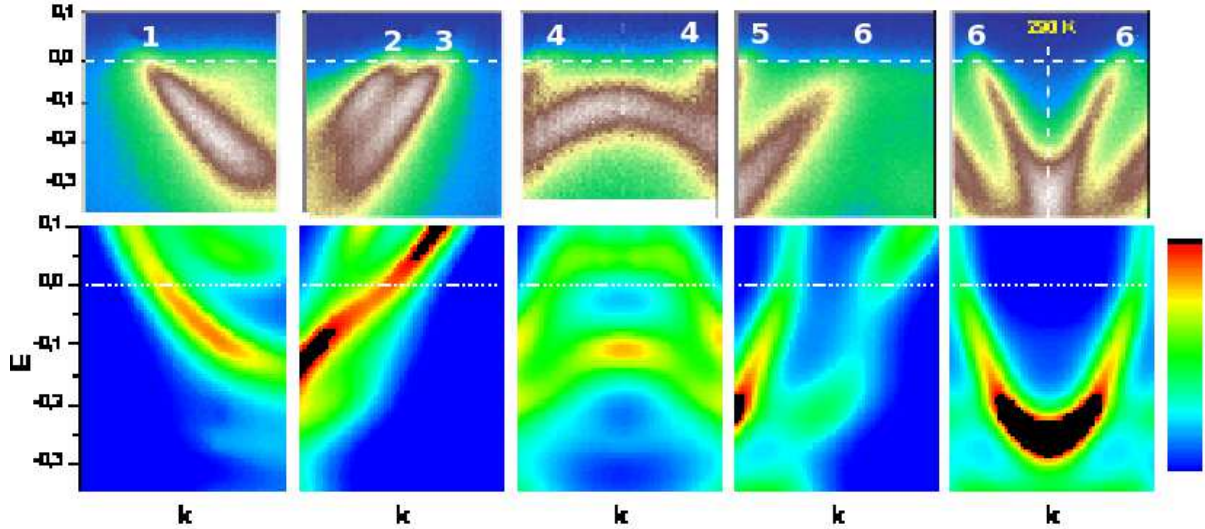}
\caption{Spectral functions for incommensurate pseudogap phase along cuts shown on Fig. 1c}
\label{c_pg}
\end{center}
\end{figure}

\begin{figure}
\begin{center}
\includegraphics[width=1.\textwidth]{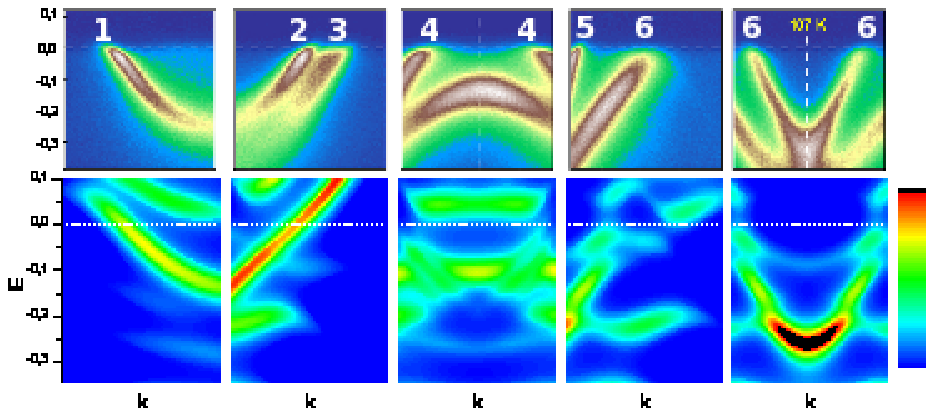}
\caption{The same as Fig. \ref{c_pg} but for incommensurate CDW phase.}
\label{c_iccdw}
\end{center}
\end{figure}

\begin{figure}
\begin{center}
\includegraphics[width=1.\textwidth]{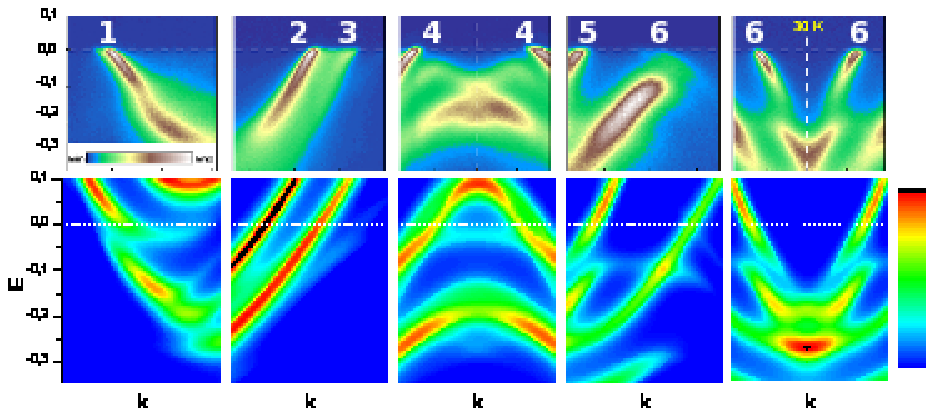}
\caption{The same as Fig. \ref{c_pg} but for commensurate CDW phase.}
\label{c_ccdw}
\end{center}
\end{figure}

Figures \ref{c_pg}--\ref{c_ccdw} show spectral functions maps in the vicinity of
the Fermi level along cuts shown on Fig. \ref{fig1}c. In all figures upper row represents
experimental data of Ref. \cite{MSe2_3}, while lower shows our theoretical results.
Generally speaking for all phases: incommensurate pseudogap Fig. \ref{c_pg},
incommensurate CDW Fig. \ref{c_iccdw} and commensurate CDW Fig. \ref{c_ccdw},
we obtain quite good qualitative agreement of theory and experiment for the number of bands 
crossing the Fermi level, their position and relative intensity.

\begin{figure}
\begin{center}
\includegraphics[width=.4\textwidth]{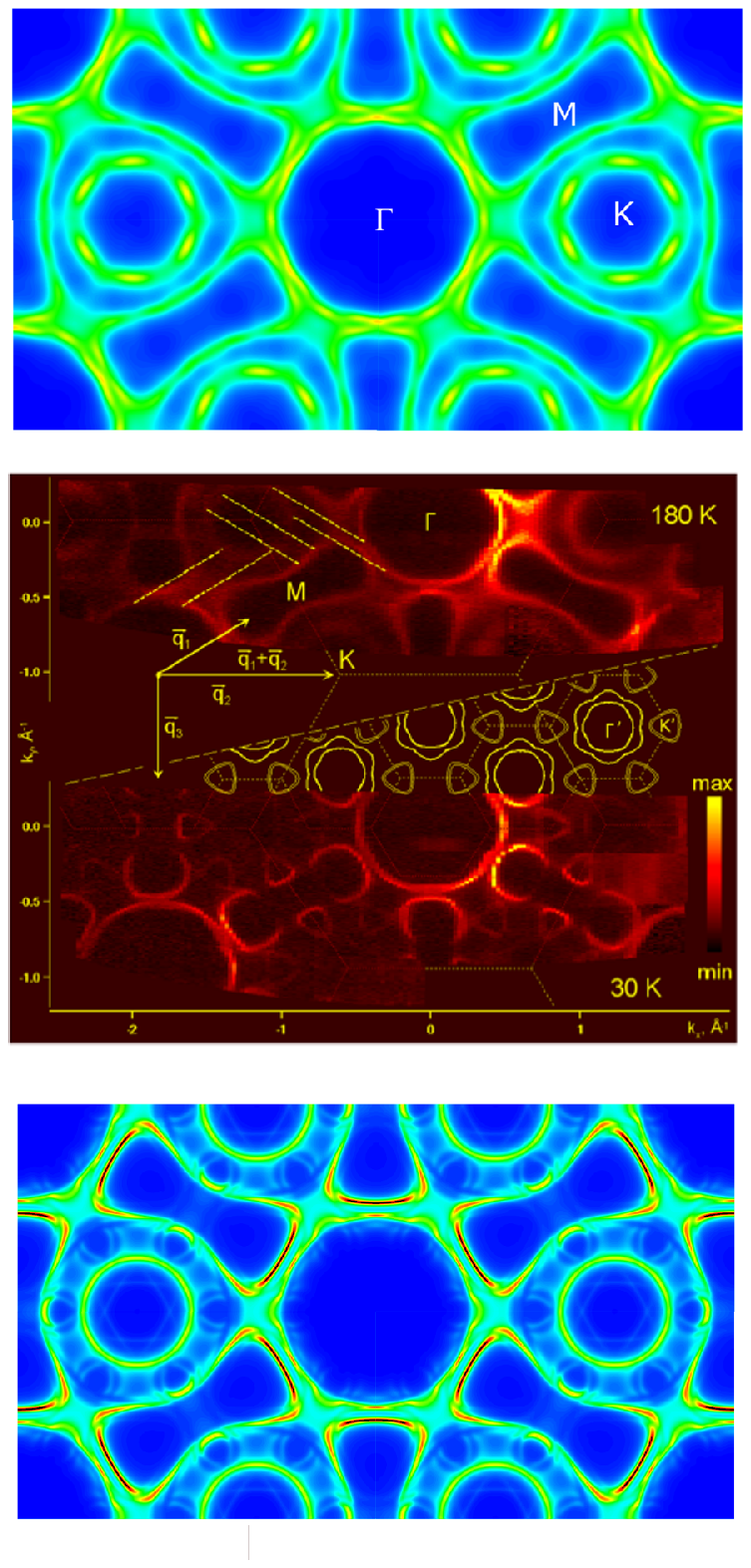}
\caption{Comparison of experimental and theoretical
fermi surfaces for 2H-TaSe$_2$.
Upper panel -- theoretical Fermi surface for pseudogap CDW phase;
middle panel -- joint picture of experimental data
pseudogap phase (upper part) and commensurate CDW phase (lower part).
Lower panel -- theoretical Fermi surface for commensurate CDW phase.
}
\label{fs_tase2}
\end{center}
\end{figure}

In Fig. \ref{fs_tase2} we present comparison of experimental and theoretical
Fermi surfaces for 2H-TaSe$_2$. In the middle part of Fig. \ref{fs_tase2} we show experimental
ARPES data from Ref. \cite{MSe2_3}.  Data at 180K corresponds to the pseudogap phase,
while those at 30K are for commensurate CDW phase. 

Upper panel of Fig. \ref{fs_tase2} shows our theoretical Fermi surface in the incommensurate 
pseudogap regime for 2H-TaSe$_2$. In general it more or less reminds
LDA Fermi surface from Fig. \ref{fig1}c. However there are obvious signatures of partial
destruction of the Fermi surface sheets. Namely, cylinder around K-point and
``bones'' along K-M direction are partially smeared out. It is seen that this picture
agrees well with the experimental ARPES data of Ref. \cite{MSe2_1,MSe2_3}.

For the commensurate CDW phase (lower panel of Fig. \ref{fs_tase2}) 
Fermi surface stays close to that obtained in LDA and shown in
of Fig. \ref{fig1}c. In contrast to incommensurate pseudogap phase
Fermi surface sheets here are more sharp both in experiment and  in theory.
The cylinder around K-point is now continuous. In the middle of the ``bones''
we observe the start of formation of small triangles as shown in the center of middle panel.
Thus, here in commensurate CDW phase of 2H-TaSe$_2$ we again obtain an overall agreement 
between theory and experiment.

\begin{figure}
\begin{center}
\includegraphics[width=.4\textwidth]{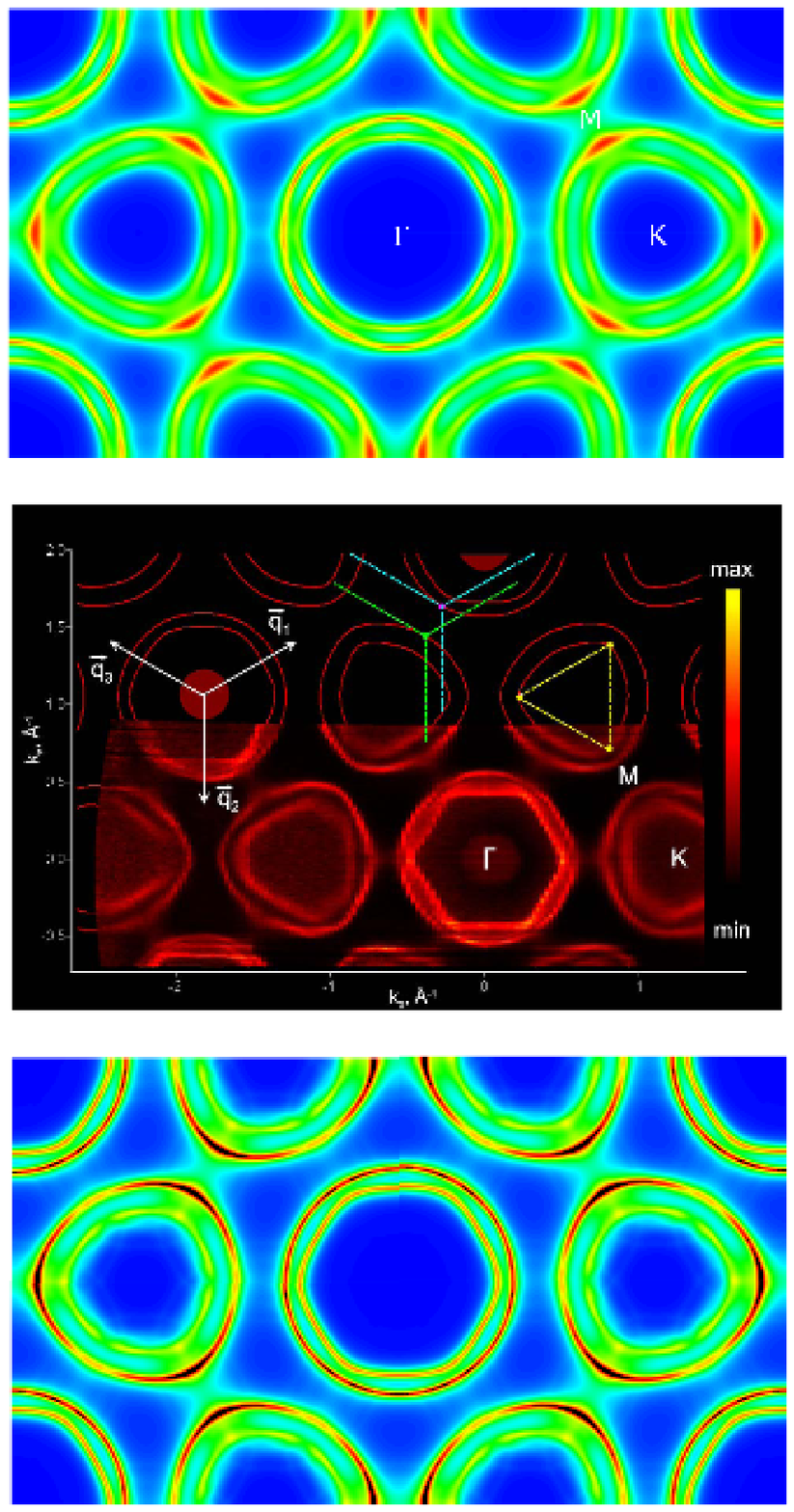}
\caption{Comparison of experimental and theoretical
Fermi surfaces for 2H-NbSe$_2$.
Upper panel -- theoretical Fermi surface for pseudogap CDW phase;
middle panel -- joint picture of experimental data
LDA Fermi surface (upper part) and commensurate CDW phase (lower part).
Lower panel -- theoretical Fermi surface for commensurate CDW phase.
}
\label{fs_nbse2}
\end{center}
\end{figure}

In Fig. \ref{fs_nbse2} we show comparison between experimental (middle panel) and theoretical
Fermi surfaces (lower and upper panel) for 2H-NbSe$_2$. 
Experimental data on the Fermi surface are available only for commensurate CDW phase \cite{MSe2_4}.
Thus we can compare these with theoretical picture shown on lower panel of Fig. \ref{fs_nbse2}.
In general,  both Fermi surfaces remind those from Fig. \ref{fig1}b. Account of electron scattering on commensurate
CDW leads to a small regions of  Fermi surface destruction, namely, along $\Gamma$-K and K-M directions.
If there exists incommensurate pseudogap phase for 2H-NbSe$_2$ at high enough temperatures,
its Fermi surface will not be changed much by pseudogap fluctuations, as seen in the upper panel of Fig. \ref{fs_nbse2}).

\section{Conclusion}
\label{concl}

To conclude, here we presented theoretical results on electronic structure of
two-dimensional diselenides 2H-TaSe$_2$ and  2H-NbSe$_2$ 
within different CDW phases.

First of all we formulated a theoretical approach to account for multiple 
scattering of electrons on different types of CDW, also for the multiple
bands case. Further we investigated spectral functions and Fermi surfaces for 
the pseudogap, incommensurate CDW and commensurate CDW phases.
Calculated theoretical spectral functions within the pseudogap phase demonstrate
``hot regions'', where spectral function is additionally broadened.
In incommensurate CDW and commensurate CDW phases in a place of these
``hot regions'' we obtained opening of the number of gaps at the intersections 
with rather pronounced ``shadow bands''. 
Comparing experimental and theoretical Fermi surfaces in the pseudogap phase
we observe rather clear signs of partial Fermi surface 
destruction with formation of a number of typical ``Fermi arcs'', separated
by pseudogap regions. In commensurate CDW phase Fermi surfaces are rather 
similar to initial LDA picture, with pretty small features due to CDW. 
The overall agreement between theory and ARPES experiments is rather 
satisfactory.  

\section{Acknowledgements}
We thank S.V. Borisenko for his interest and helpful discussions.
This work is partly supported by RFBR grant 11-02-00147 and was performed
within the framework of Programs of Fundamental Research of the Russian 
Academy of Sciences (RAS) ``Quantum physics of condensed matter'' 
(UB RAS 09-$\Pi$-2-1009) and of the Physics Division of RAS  ``Strongly correlated 
electrons in solid states'' (UB RAS 09-T-2-1011). 

\newpage

\appendix

\section{Appendix: Scattering on commensurate CDW}

\subsection{One band scattering}


First let us consider single band case with electronic ``bare'' spectrum 
$\epsilon ({\bf k})$. These ``bare'' electrons are scattered on CDW 
potential, written as:
\begin{equation}
V({\bf r})=2\Delta \sum_{l=1}^{3}\mathrm{cos}{\bf Q}_l{\bf r}.
\label{CDWpot}
\end{equation}
``Bare'' retarded Green's function is:
\begin{equation}
g({\bf k})=\frac{1}{E-\epsilon ({\bf k})+i\delta}.
\label{G0}
\end{equation}
Let us introduce short notations: $g({\bf k})=g$; $g({\bf k}+{\bf Q}_l)=f_l$; 
$g({\bf k}-{\bf Q}_l)=f_{\bar l}$; $g({\bf k}+{\bf X}_l)=\phi_ l$; $g({\bf k}-{\bf X}_l)=\phi_{\bar l}$. 
Then for diagonal Green's function $G=G({\bf k},{\bf k})$ and twelve off-diagonal 
($F_l=G({\bf k}+{\bf Q}_l,{\bf k})$; $F_{\bar l}=G({\bf k}-{\bf Q}_l,{\bf k})$; 
$\Phi_l=G({\bf k}+{\bf X}_l,{\bf k})$; $\Phi_{\bar l}=G({\bf k}-{\bf X}_l,{\bf k})$) 
one can get the following system of thirteen linear equations
(see Table I of scattering vectors summation):
\begin{equation}
\begin{split}
G&=g+g\Delta F\\
F_1&=f_1\Delta (F_{\bar 1}+\Phi_2+F_2+G+F_{\bar 3}+\Phi_1)\\
F_2&=f_2\Delta (\Phi_2+F_{\bar 2}+\Phi_3+F_3+G+F_1)\\
F_3&=f_3\Delta (F_2+\Phi_3+F_{\bar 3}+\Phi_{\bar 1}+F_{\bar 1}+G)\\
F_{\bar 1}&=f_{\bar 1}\Delta (G+F_3+\Phi_{\bar 1}+F_1+\Phi_{\bar 2}+F_{\bar 2})\\
F_{\bar 2}&=f_{\bar 2}\Delta (F_{\bar 3}+G+F_{\bar 1}+\Phi_{\bar 2}+F_2+\Phi_{\bar 3})\\
F_{\bar 3}&=f_{\bar 3}\Delta (\Phi_1+F_1+G+F_{\bar 2}+\Phi_{\bar 3}+F_3)\\
\Phi_l&=\phi_l\Delta F; \qquad \Phi_{\bar l}=\phi_{\bar l}\Delta F,
\end{split}
\label{system}
\end{equation}
where $F=\sum_{l=1}^{3}(F_l+F_{\bar l})$. 

Solving Eqs. (\ref{system}), one can the get diagonal
Green's function $G=G({\bf k},{\bf k})$:
\begin{equation}
G=gK; \qquad K=\frac{1-\alpha\beta-a(\beta +1)-b(\alpha +1)}
{1-\alpha\beta-a(\beta +1)-b(\alpha +1)-g\Delta [\alpha (\beta +1)+\beta (\alpha +1)]},
\label{G_1z}
\end{equation}
where $\alpha =\Delta (f_2+f_{\bar 1}+f_{\bar 3})$, $\beta =\Delta (f_1+f_3+f_{\bar 2})$, 
$a=\Delta^2[f_{\bar 1}(\phi_{\bar 1}+\phi_{\bar 2})+f_2(\phi_2+\phi_3)+f_{\bar 3}(\phi_{\bar 3}+\phi_1)]$, 
$b=\Delta^2[f_1(\phi_1+\phi_2)+f_3(\phi_3+\phi_{\bar 1})+f_{\bar 2}(\phi_{\bar 2}+\phi_{\bar 3})]$.

\subsection{Multiband scattering}


Following the approach of Ref. \cite{PG_FeAs} we assume that CDW scattering 
amplitude $\Delta$ is identical for intra and interband transitions. 
One can define short notations:
$g^i=g^i({\bf k})=\frac{1}{E-\epsilon_i ({\bf k})+i\delta}$; $f^i_{l(\bar l)}=g^i({\bf k}\pm{\bf Q}_l)$; 
$\phi^i_{l(\bar l)}=g^i({\bf k}\pm{\bf X}_l)$,
where $i$ -- band index. Diagonal and off-diagonal Green's functions will have 
additional band indices
The rest of notations are the same as for single band case.
For diagonal Green's function, in analogy with first equation of the system
(\ref{system}), one can obtain:
\begin{equation}
G^{ij}=g^i\delta_{ij}+g^i\Delta \sum_{m}\sum_{l=1}^{3}(F_l^{mj}+F_{\bar l}^{mj}).
\label{Gdiag}
\end{equation}
Introducing $G^j=\sum_iG^{ij}$; $F_{l(\bar l)}^j=\sum_iF_{l(\bar l)}^{ij}$; 
$\Phi_{l(\bar l)}^j=\sum_i\Phi_{l(\bar l)}^{ij}$; $g=\sum_ig^i$; $f_{l(\bar l)}=\sum_if_{l(\bar l)}^i$; 
$\phi_{l(\bar l)}=\sum_i\phi_{l(\bar l)}^i$ and summing Eq. (\ref{Gdiag}) 
over $i$ we get: 
\begin{equation}
G^j=g^j+g\Delta \sum_{l=1}^{3}(F_l^j+F_{\bar l}^j).
\label{Gj}
\end{equation}
The rest of other twelve equations for $F_{l(\bar l)}^j$ and $\Phi_{l(\bar l)}^j$
are completely equivalent to corresponding equations of one band case (\ref{system}).
Thus, we immediately obtain:
\begin{equation}
G^j=g^jK,
\label{G_nz}
\end{equation}
where $K$ is defined in Eq. (\ref{G_1z}). However, now the 
quantities $g$, $f_{l(\bar l)}$, $\phi_{l(\bar l)}$ 
are summed up over all band indices. From Eq. (\ref{Gdiag}) and (\ref{Gj}), 
using Eq. (\ref{G_nz}) we finally obtain: 
\begin{equation}
G^{ij}=g^i\delta_{ij}+g^i\frac{G^j-g^j}{g}=g^i\delta_{ij}+\frac{g^ig^j}{g}(K-1),
\label{Gij_nz}
\end{equation}
which allows us to calculate the spectral function (\ref{SpDen}) with the 
account of scattering on commensurate CDW.

\newpage

\newpage

\newpage

\newpage

\newpage

\newpage

\newpage

\newpage

\newpage

\newpage

\newpage

\newpage

\newpage


\begin{thebibliography}{99}

\vfill
\bibitem{Wilson74}J.A. Wilson, F.J. Di Salvo, S. Mahajan, Adv. Phys. {\bf 24}, 117 (1975).

\bibitem{Moncton} D.E. Moncton, J.D. Axe, and F.J. Di Salvo, Phys. Rev. Lett. {\bf 34}, 734 (1975);
Phys. Rev. B {\bf 16}, 801 (1977).

\bibitem{MSe2_1} D. S. Inosov, V. B. Zabolotnyy, D. V. Evtushinsky, A. A. Kordyuk,
B. Buchner, R. Follath, H. Berger, and S. V. Borisenko New J. Phys. 10, 125027 (2008). 

\bibitem{MSe2_3}S. V. Borisenko , A. A. Kordyuk, A. N. Yaresko, V. B. Zabolotnyy, D. S. Inosov, 
R. Schuster, B. Buchner, R. Weber, R. Follath, L. Patthey, H. Berger. 
Phys. Rev. Lett. {\bf 100}, 196402 (2008).

\bibitem{MSe2_4}S. V. Borisenko, A. A. Kordyuk, V. B. Zabolotnyy, D. S. Inosov, D. Evtushinsky, 
B. Buchner, A. N. Yaresko, A. Varykhalov, R. Follath, W. Eberhardt, L. Patthey, H. Berger. 
Phys. Rev. Lett. {\bf 102}, 166402 (2009)

\bibitem{LMTO}O.K. Andersen. Phys. Rev. B {\bf 12}, 3060 (1975);
O. Gunnarsson, O. Jepsen,  O.K. Andersen. Phys. Rev. B {\bf 27}, 7144 (1983);
O.K. Andersen, O. Jepsen.  Phys. Rev. Lett. {\bf 53}, 2571  (1984).

\bibitem{elstruc} R.A. Bromley, Phys. Rev. Lett. {\bf 29}, 357 (1972);
L.F. Mattheiss, Phys. Rev. B {\bf 8}, 3719 (1973);
G. Wexler, A.M. Wooley J. Phys. C: Solid State Phys. {\bf 9}, 1185 (1976);
R. Corcoran $et~al.$, J. Phys.: Condens. Matter {\bf 6}, 4479 (1994);
H.E. Brauer $et~al.$, J. Phys.: Condens. Matter {\bf 13}, 9879 (2001);
M.-T. Suzuki, H.Harima, Physica B, {\bf 359-361}, 1180 (2004).

\bibitem{Smith} N.V. Smith, S.D. Kevan, F.J. Di Salvo, J. Phys. C: Solid State Phys. {\bf 18}, 3175 (1985).

\bibitem{Rossnagel} K. Rossnagel, E. Rotenberg, H. Koh, N.V. Smith, and L. Kipp,  Phys. Rev. B {\bf 72}, 121103 (2005).

\bibitem{MSe2_2}D. S. Inosov, D. V. Evtushinsky, V. B. Zabolotnyy, A. A. Kordyuk, B. Buchner, 
R. Follath, H. Berger, S. V. Borisenko. 
Phys. Rev. B {\bf 79}, 125112 (2009).


\bibitem{PG_FeAs}E.Z.Kuchinskii, M.V.Sadovskii.
Pis'ma v ZhETF {\bf 91}, 729 (2010) [JETP Lett. {\bf 91}, 660 (2010)]. 

\bibitem{Sad00}M.V.Sadovskii.  
Zh. Eksp. Teor. Fiz. {\bf 66}, 1720 (1974) [Sov. Phys. - JETP {\bf 39}, 845 (1974)];
Zh. Eksp. Teor. Fiz. {\bf 77}, 2070 (1979) [Sov. Phys. - JETP {\bf 50}, 989 (1979)].

\bibitem{KS99}E.Z. Kuchinskii, M.V. Sadovskii,
Zh. Eksp. Teor. Fiz. {\bf 115}, 1765 (1999) [JETP {\bf 88}, 968 (1999)]. 

\bibitem{Sch} J. Schmalian, D. Pines, B. Stojkovic. 
Phys. Rev. Lett. {\bf 80}, 3839(1998); 
Phys. Rev. B {\bf 60}, 667 (1999).

\bibitem{MS}M.V. Sadovskii, Usp. Fiz. Nauk {\bf 171}, 539 (2001) [Physics-Uspekhi {\bf 44}, 515 (2001)];\ 

Also in {\it Strings, branes, lattices, networks, pseudogaps and dust} 
(Moscow:Scientific World, 2007) p. 357 (in Russian, English version --
ArXiv:\ cond-mat/0408489).

\end{thebibliography}
\end{document}